\documentclass[twocolumn,showpacs,preprintnumbers,amsmath,amssymb]{revtex4}

\usepackage{graphicx}
\usepackage{bm}
\usepackage{color}

\newcommand{\V }[1]{\overrightarrow{#1}}

\newcommand{\be}{\begin{equation}}
\newcommand{\ee}{\end{equation}}
\newcommand{\bee}{\begin{eqnarray}}
\newcommand{\eee}{\end{eqnarray}}
\newcommand{\shear}{\gamma}

\newcommand{\sigP}{\sigma_{yy}}
\newcommand{\sigD}{\sigma_{xy}}
\newcommand{\young}{E}

\newcommand{\viscofluid}{\eta}

\newcommand{\gap}{h}
\newcommand{\tstokes}{t_{St}}

\begin{document}

\preprint{APS/123-QED}

\title{Internal relaxation time in immersed particulate materials.}

\author{P. Rognon}
\author{I. Einav}
\email{i.einav@usyd.edu.au}
\affiliation{
School of Civil Engineering, J05, The University of Sydney, Sydney, New South Wales 2006,
Australia.}
\author{C. Gay}
\email{cyprien.gay@univ-paris-diderot.fr}

\affiliation{Mati\`{e}re et Syst\`{e}mes Complexes, 
 Universit\'{e} Paris Diderot--Paris 7, CNRS, UMR~7057, 
 B\^atiment Condorcet, Case courrier 7056, F--75205 PARIS cedex 13, France.
}


\date{\today}
\begin{abstract}
We study the dynamics of the solid to liquid transition for a model material made of
elastic particles immersed in a viscous fluid. The interaction between
particle surfaces includes their viscous lubrication, a sharp repulsion when they get
closer than a tuned steric length and their elastic deflection induced by those two
forces. We use {\em Soft Dynamics} to simulate the dynamics of this material when
it experiences a step increase in the shear stress and a constant normal stress.
We observe a long creep phase before a substantial flow eventually establishes. 
We find that the typical creep time relies on an internal relaxation process,
namely the separation of two particles driven by the applied stress 
and resisted by the viscous friction.
This mechanism should be relevant for granular pastes, living cells,
emulsions and wet foams.
\end{abstract}


\maketitle

\section{Introduction}

The rheological behavior of soft amorphous materials such as
foams~\cite{Hohler05,Denkov09SM}, emulsions~\cite{Becu06,Derkach09,Bocquet09}, living
cells~\cite{Yilmaz08} and granular pastes~\cite{Stickel05,Coussot05} is an active field of
research with many practical issues, for these materials are broadly
involved in
geophysical events, industrial processes and biological organisms. Besides, understanding
their dynamics presents great theoretical challenges as their structure, similar to that
of glasses, is disordered, metastable and out of thermodynamical
equilibrium~\cite{Sollich97,Richard05}. 

Perhaps the most salient property of these materials is their ability to deform
like a solid or to flow like a liquid depending on the loading conditions. At the
microscopic level, this solid-liquid transition relies on the particles ability to move
with respect to each other, similar to the cage effect in molecular or colloidal
systems~\cite{Marty05,Coussot07}.
However, large enough bubbles, droplets and grains do not show significant Brownian
motion and their solid-liquid transition is thus not thermally activated. The
\textit{driving force} of the relative particle motion comes only from the external
loading, while the \textit{resisting force} comes from the particle interactions and/or deformations. This
force balance defines internal dynamics with an associated
time-scale, which is the core of the thixotropy
effects in
various materials~\cite{Barnes97,Dacruz02,Moller06,Coussot07,Grillet09} and of the
relaxation time evidenced by oscillatory shear experiments~\cite{Narumi05,Wyss07}. The
interstitial liquid is pivotal to these internal dynamics, as it affects both the
\textit{driving force} and the \textit{resisting force} that the particles experience.

Immersed particulate materials support the stresses
not only through their contact
network (known as osmotic or effective stresses), but also by the continuous liquid (pore
pressure). This
notion of pore pressure is a central point of soil stability analysis
(see~\cite{Iverson05} and reference therein), and was shown to be responsible for a
long delay in the triggering of submarine granular avalanches~\cite{Pailha08}. The
underlying mechanism involves the dilatancy needed to start shearing and the time
required to fill the growing pores with liquid. This filling time then relies on the
degree of drainage of the shearing zone, in other word on how long the liquid takes to
flow from one part of the material to another. Assuming that the granular skeleton does not deform, 
the flow of liquid through the pore network satisfies
a Darcy law with some permeability~\cite{Pailha08}.

Besides the pore pressure, the liquid has a strong effect 
on the particle interactions as it lubricates close pairs of surfaces. 
The viscous friction between two smooth rigid spheres is a
well known problem (see for instance~\cite{Batchelor67}). However, the presence of
asperities on rough surfaces has been shown to strongly affect lubrication, and
this topic is still an active field of research
\cite{Jenkins05,Vinogradova06,Yang06JFM,Divoux07}. Between bubbles or droplets, the flow
of liquid is even more complex, as the surfaces are soft and can
stretch~\cite{Lacasse96a,Durand06}, they can be permeable~\cite{Sagis07} and exhibit some
repulsive electrostatic interactions leading to disjoncting pressure~\cite{Tabakova09} as
well as adhesion~\cite{Besson08,Denkov09SM,Denkov09PRL}.

This paper presents a numerical investigation on the dynamics of the solid-liquid
transition in a model immersed particulate material. Our motivation is to capture the
common characteristics
of bubbles, droplets or solid grains: all of these particles can deform elastically
and interact through lubricated contacts. Therefore, we do not address the flow of
the liquid
in the pore network or any of the others interactions mentioned
above. Instead, here we focus on elastic spheres interacting \textit{via}
lubricated contacts with a steric repulsion when surfaces get very close. We
simulate the dynamics of this material using {\em Soft Dynamics}, the discrete
element method that we introduced in Refs.~\cite{Rognon08EPJE,Rognon09EPJE}. 
We subject large samples to a constant normal (osmotic) stress and a (smoothed)
stepwise increase of the shear stress. 
We then measure the delay before the onset of a steady material flow.
Finally, we show that this macroscopic delay reflects
the time taken by two particles for separating, 
a motion driven by a tensile force
that results from the applied stress
and that is resisted by the normal viscous friction.

\section{Internal dynamics: insights from dimensional analysis}\label{sec:DA}

Dimensional analysis is often used to define the typical basic times that
the system involves. Let us recall its outcome, as expressed in various
studies. We are dealing with elastic particles of size $d$, Young's modulus $\young$
and
mass $m$. They are immersed in a Newtonian fluid (viscosity $\viscofluid$) and
confined
by an osmotic normal stress $\sigP$. Four time scales emerge from those quantities.

The first two time scales correspond to the motion of a single particle which experiences
a driving force $\sigP d^2$. For dry grains, this motion is resisted only by the grain
inertia (mass $m$), while for particles with negligible mass 
immersed in a fluid of viscosity $\viscofluid$, it is resisted merely by the Stokes drag.
The respective corresponding times 
are the \textit{inertial time} $t_{i}$~\cite{Dacruz05b,GDR04,Forterre08} 
and the \textit{Stokes time} $\tstokes$~\cite{Cassar05,Doppler07}:
\bee
t_{i}&=&\sqrt{\frac{m}{\sigP d}} ; \label{eq_ti}\\
\tstokes &=& 3 \pi \frac{\viscofluid}{\sigP}.\label{eq_tst}
\eee

The other two time scales can be directly obtained by replacing the scale of stress
$\sigP$ by the Young modulus $\young$ in each above time scale.
This leads
to the time $t_c$ for sound to propagate through a grain 
(which can also be linked to the binary collision time~\cite{Campbell05})
and to the time $t_{M}$ for an elastically deformed particle to recover its shape when immersed in a viscous
fluid, a time similar to that of a Maxwell relaxation~\cite{Duran95,Duran97}:
\bee
t_{c}&=&\sqrt{\frac{m}{\young
d}} \label{eq_tc}\\
t_{M}&=& \frac{\viscofluid}{\young}\label{eq_tM}
\eee

Tab.~\ref{tab:times} gives an estimate of those time scales
for elastic grains and for bubbles. 
For the particle deformation to remain moderate,
the reduced normal stress $\kappa=\frac{\sigP}{\young}$ must be small, 
and time scale $t_{c}$ (resp. $t_{M}$) is thus 
shorter than $t_{i}$ (resp. $\tstokes$).

The above references clearly show the usefulness of such time scales,
especially for expressing flow laws.
However, none of them accounts for the details of the interaction
of a particle with its neighbourhood. 
The aim of the present paper is to point out the importance 
of such a time scale when particles are elastic
and interact {\em via} lubricated contacts.

\begin{table}
\caption{Order of magnitudes of the internal time scales in immersed particulate systems.
Time scales from dimensional analysis (defined in
section~\ref{sec:DA}) and delay time scale $\Delta$ from lubrication interactions
(computed through the present study). 
Particles consist in grains or bubbles of size $d=1$~mm
immersed in water (viscosity $10^{-3}$~Pa\,s). 
The grain density is
$2.10^3$~kg\,m$^{-3}$ and their Young modulus is $10^{10}$~Pa. The density of bubbles is
null and their effective Young modulus is of the order of $\young \approx
10^{2}$~Pa. We consider an osmotic normal stress of $10^7$ Pa for the
grains and $10^{-1}$~Pa for the bubbles 
in order to obtain the same reduced normal stress 
$\kappa = \sigP/\young = 10^{-3}$ in both cases. 
The delay $\Delta$ is estimated according to Eq.(\ref{eq:tsep}) with a steric
length of $h_0=10^{-7}$~m, which corresponds to particles in the Hertz limit.}
\label{tab:times}   
 \begin{ruledtabular}
\begin{tabular}{lccccc}
   Time (s)    & $t_{i}$ & $\tstokes$ & $t_{c}$ & $t_{M}$ & $\Delta$ \\ 
\hline\noalign{\smallskip}
Grains & $10^{-5}$ & $10^{-9}$ & 
$10^{-7.5}$ & $10^{-13}$ & $10^4 \tstokes$  \\ 
\hline\noalign{\smallskip}
Bubbles & $0$ & $10^{-1}$ & $0$ &  $10^{-5}$
& \footnote{The above delay $\Delta$ was
based on Hertzian elasticity.
Of course, bubbles and droplets do not display any bulk elasticity,
yet they do recover their shape upon load reversal.
For the present purpose, their effective Young modulus
was estimated as the surface tension ($10^{-1}$~N\,m$^{-1}$) 
divided by the bubble size: $\young \approx 10^{2}$~Pa 
(see~\cite{Lacasse96a} for more details on droplet contact).}$10^4 \tstokes$
\end{tabular}
\end{ruledtabular}
\end{table}

\begin{figure*}[!htb]
\begin{center}
\resizebox{1.9\columnwidth}{!}{%
\includegraphics{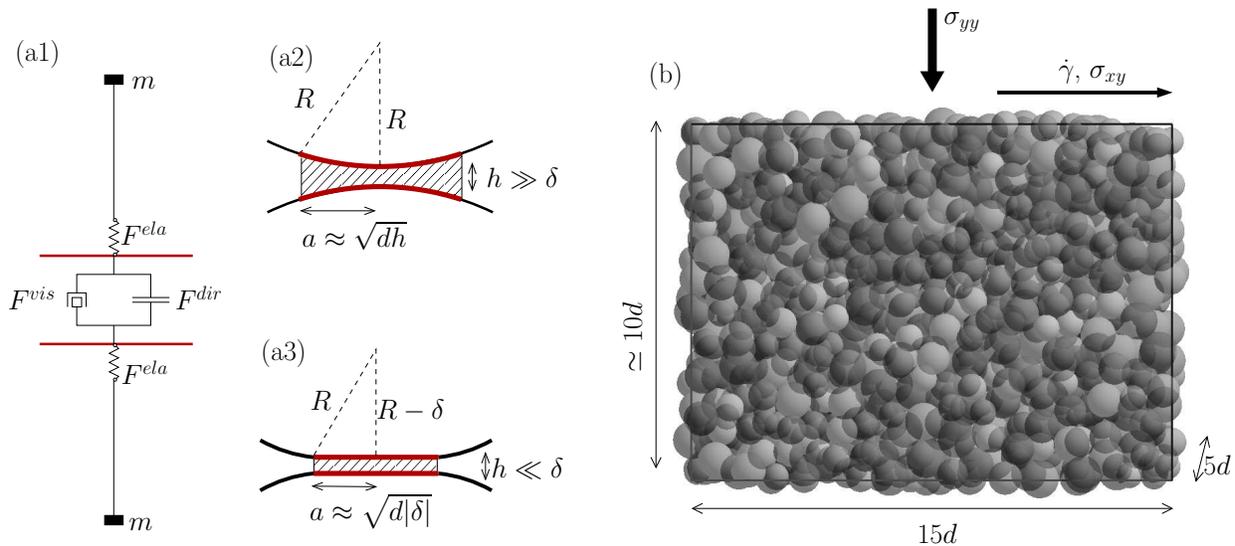}
}
\end{center}
\caption{(Color online) {\em (a)} Particle-fluid-particle interaction: 
{\em (a1)} the force $F^{ela}$ that deforms the surface elastically
decomposes into the viscous force  $F^{vis}$ and the direct particle-particle
force $F^{dir}$.
The size $a$ of the transmitting force region 
is represented in two asymptotic limits: 
{\em (a2)} the Poiseuille limit for non-deflecting spheres, and 
{\em (a3)} the Hertz limit when the deflection $\delta$ is much larger 
than the gap width $\gap$.
{\em (b)} Plane shear simulated using {\em Soft Dynamics}. 
1000 particles are subjected to normal stress $\sigP$ and shear stress $\sigD$ 
in a periodic domain (Lees-Edwards boundary conditions~\cite{Lees72}). 
The height of the cell fluctuates freely
in order to keep $\sigP$ constant.
}
\label{Fig1}
\end{figure*}

\section{Internal dynamics: particle interaction}~\label{sec:interaction}

We consider an ideal suspension made of elastic particles in a Newtonian
fluid. We focus on dense configurations, where the surface-to-surface distance $h$
between a particle and its few closest neighbours is typically much smaller than the
particle diameter $d$. 
Under such conditions, the particle-fluid-particle interaction is mostly governed by the
liquid that is present within a small gap of thickness $h$ and radius $a$ with $h \ll a \ll d$ (Fig.\ref{Fig1}a).
Accordingly, the force and torque transmitted from surface to surface
within each pair of neighbouring particles
are proportional to their relative velocity. 
For such a gap with a low aspect ratio ($\gap\ll a$)
in which the flow satisfies the conditions 
of the lubrication approximation,
the proportionality coefficients
(which we here call viscous frictions)
can be expressed rather easily 
(see Appendix~\ref{app:friction})
as functions of the gap thickness $\gap$ and size $a$.

As a result of the viscous force and torque
transmitted through the gap,
the particle surfaces locally deform elastically. 
Hence, for each pair of close neighbours,
the combination of the gap viscous response and the particle elasticities
is analogous to a Maxwell element (see Fig.\ref{Fig1}, part a1)
in which both the stiffness and the viscous friction
are non-linear (in the sense that they depend on the gap geometry)
and tensorial (since they reflect the normal,
tangential and twist modes, see Fig.~\ref{Fig5}). 

The viscous friction term that represents the normal relative
motion (Poiseuille flow) turns out to be typically
much higher than the corresponding terms
for the tangential sliding mode (Couette flow) and the twist mode. 
Thus, the normal relative motion 
should be responsible for the longest relaxation time. 
Let us therefore recall the expression of the normal viscous
and elastic forces (see Appendix~\ref{app:friction}):

\begin{eqnarray} 
 F^{vis}_n &=& \frac{3}{2} \pi \viscofluid \frac{a^4}{\gap^3}\dot \gap
\label{Eq:viscous_force} \\
F^{ela}_n &=& a\young \delta \label{Eq:elastic_force}
\end{eqnarray}

\noindent where $\young$ is the Young modulus of the particle 
and $\delta$ the normal surface deflection. 
The size $a$ of the region that transmits most of the force
basically depends on the distance between surfaces 
and on the flattening induced by their elastic deflections.
If the elastic deflection is much smaller than the gap thickness, 
the surface shape is mostly spherical
and the size $a$ is given by the classical 
lubrication approximation between two spheres:
$a \approx \sqrt{dh}$ (Fig.\ref{Fig1} a2).  
We refer to this asymptotic limit as the \textit{Poiseuille} limit.
Conversely, if the elastic deflection 
is much larger than the gap thickness, the
particle surface elastically flattens 
over a region whose size is given by Hertz's theory: 
$a\approx \sqrt{d\delta}$ (Fig.\ref{Fig1} a3). 
We refer to this asymptotic limit as the
\textit{Hertz} limit. 

So far, two particles could not statically sustain 
any nonzero confining force, 
since the fluid would keep flowing out of the gap 
and the particle surfaces would correspondingly 
keep approaching although in a slower and slower motion. 
In practice, however, there is usually a typical distance
$\gap_0$ below which particle
surfaces interact not only \textit{via} the fluid, 
but also through a direct particle-particle force. 
This interaction can have various physical origins, 
among which direct contacts between asperities of
rough solid surfaces~\cite{Jenkins05,Vinogradova06,Yang06JFM,Divoux07} 
or repulsion between charged or polymeric surfactant molecules
at the bubble or droplet surface~\cite{Tabakova09}.
Our present focus is not necessarily to capture 
any of the details of such effects, 
but rather to represent their common feature: 
a strong steric repulsion that works on the particle surfaces 
in parallel to the viscous force (Fig.\ref{Fig1} a1). 
For that purpose, we chose the following power law: 

\begin{equation}
F^{dir}_n = F^{conf}_n \,
\left( \frac{\gap}{\gap_0}
\right)^{-\alpha}.
\end{equation}

\noindent 
For $\alpha\gg 1$, this law captures our requirement 
for a strongly increasing repulsion as the gap becomes thinner 
than the steric length $h_0$, and a quickly vanishing repulsion for
thicker gaps. Ideally, we would have liked to have an infinite $\alpha$, 
but as this function becomes more strongly non-linear,
the critical time step for accurate simulations increases tremendously.
A value of $\alpha=8$ was found as a good compromise 
between these two requirements. 
The force scale parameter is chosen as $F^{conf}_n=\sigP\,d^2$.
In this way, the surface-to-surface distances 
stabilize around $h_0$ when the material is subjected
to a confining stress $\sigP$.

\section{Collective response to a step shear stress} \label{sec:result}

In this section, we use {\em Soft Dynamics} to study the response of the
material to a sudden increase in shear stress
in plane shear geometry, in the absence of gravity, with a fully periodic domain
(Fig.\ref{Fig1}b). The principle of the method is detailed in Appendix~\ref{sec:motion}.
The typical experiment consists of two steps: {\em (i)} the preparation of a dense
compressed sample and {\em (ii)} the shear by itself. 

First, particles are set in a loose and random configuration. 
They are then subjected solely to a designated 
target normal stress $\sigP$.
This makes the material denser, with closer particles, 
while the height of the periodic cell decreases.
The material eventually reaches a static state 
where the confining stress is supported 
by some pairs of particles
whose surface to surface distances are 
on the order of the steric length $\gap_0$. 
A correct simulation of this compression 
should include the long range multibody hydrodynamics
interaction, while the particles are still in a loose configuration. 
However, we do not focus on the compression process
and use it only as a way to obtain a close-packed configuration. 
In the second step, which is the prime focus of our paper, 
we subject the configuration 
to a (smoothened) step in shear stress
from zero to a final value $\sigD$.
We then measure the time evolution
of the nominal shear strain $\shear$. 
More preciseley, the shear stress is continuously increased 
from zero to its final value within a short but finite time 
of the order of $\tstokes$ (Eq.\ref{eq_tst}) as $\sigD(1-e^{-\frac{t}{\beta
\tstokes}})$ (see Fig.~\ref{Fig2}). 
Most of the results are obtained with $\beta=1$.
Different values ($\beta=0.5$ and $\beta=2$)
will be used to probe the influence
of the rate of the initial shear stress increase.

We perform a series of such numerical experiments 
for materials with various steric lengths
$\gap_0$, subjected to various
levels of shear stress $\sigD$. 
The parameters of the system are summarized in Tab.~\ref{tab1}, 
expressed in units of particle size $d$ for the length,
normal stress $\sigP$ for the stress and Stokes time $\tstokes$ for the time. The grains have a mass,
but we set the inertial time much smaller than the Stokes time, so that the dynamics
of the system is mostly driven by viscous interactions. 
The reduced normal stress $\kappa=\sigP/\young$ 
applied to the particles
is much smaller than unity so that the elastic
deformations would be much smaller than the particle size. 
We complete the current section by
describing the resulting collective behavior 
in our {\em Soft Dynamics} simulations, 
which we will then analyse in terms 
of the internal dynamics in Section~\ref{sec:discussion}.

\begin{table}
\caption{Properties of the simulated systems: inertial time $t_i$, 
reduced applied normal stress $\kappa$, 
steric distance $h_0$ and shear stress $\sigD$.}
\label{tab1}   
 \begin{ruledtabular}
\begin{tabular}{cccc}
\hline\noalign{\smallskip}
$\frac{t_i}{\tstokes}$ & $\kappa = \frac{\sigP}{E}$ &$\frac{h_0}{d}$ &
$\frac{\sigD}{\sigP}$\\ 
\noalign{\smallskip}\hline\noalign{\smallskip}
$0.1$  & $10^{-3}$ & $8.10^{-4} \rightarrow 10^{-1}$&$ 0.05\rightarrow 1$\\
\noalign{\smallskip}\hline
\end{tabular}
\end{ruledtabular}
\end{table}

\subsection{The role of the steric length}

\begin{figure*}[!htb]
 \begin{center}
 \resizebox{2.\columnwidth}{!}{%
 \includegraphics{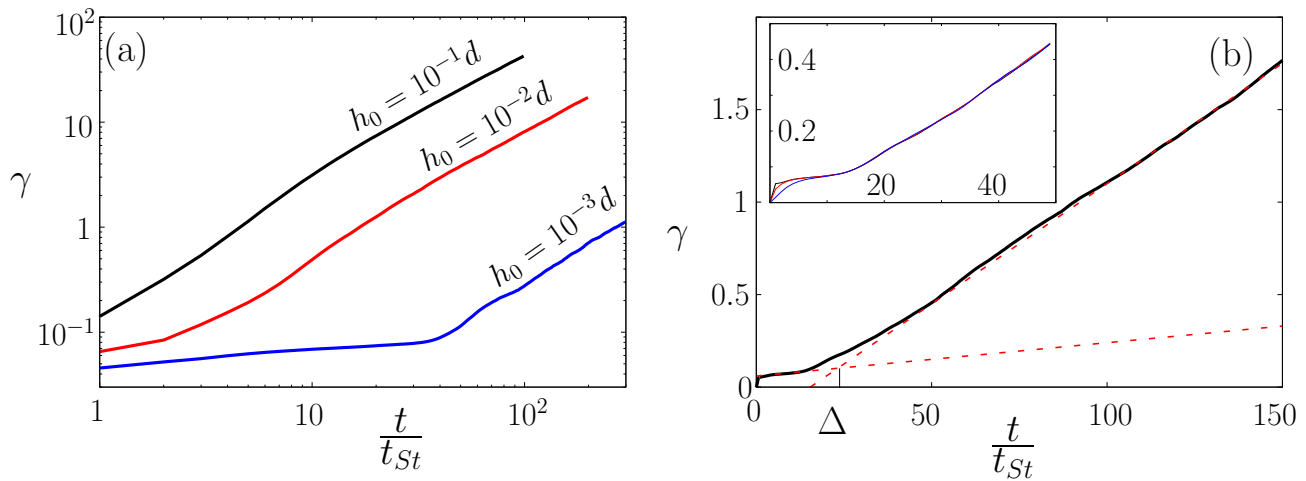}
 }
 \end{center}
 \caption{ (Color online) Time evolution of the shear strain $\shear$ induced by a rapid
increase in shear stress $\sigD$ from $0$ to 
$0.5\sigP$ between $t=0$ and $t=\tstokes$. 
{\em (a)} Materials with
a different value of the steric length $h_0$. 
{\em (b)} Procedure to determine the typical duration $\Delta$ 
of the creeping phase, here for $h_0/d =2.10^{-3}$; 
the dashed lines correspond to the
linear fit of $\shear(t)$ in the creep phase and in the flow phase;
(Inset) for the same matrial, different rates of initial stress increase: $\beta = 0.1$
(black), $0.5$ (red) and $2$ (blue) (see introduction of Section~\ref{sec:result}).
}
\label{Fig2}
\end{figure*}

In the first test we subject materials 
with different steric lengths $h_0$ 
to a similar constant shear stress $\sigD =0.5\,\sigP$ (Fig.~\ref{Fig2}a).
For the longest steric distance, $h_0/d=10^{-1}$, 
the shear strain increases almost steadily. 
In contrast, for the shortest distance $h_0/d=10^{-3}$, 
the material first deforms very slowly.
After this creep phase, it starts to flow steadily. 
During both the creep and the flow phases,
$\shear(t)$ is mostly linear (Fig.~\ref{Fig2}b).
Except for the very begining of the creeping phase, this behavior is not sensitive to the
typical time of increase of the stress (Fig.~\ref{Fig2}b, inset).
We define the typical duration $\Delta$ of
the creep phase using a geometrical construction 
that searches for the intersection
between these two lines, as plotted in linear coordinates. 
Note that prior to the creep phase, 
there is another very short phase, 
during which the shear deformation quickly increases 
up to a small level. 
This corresponds to the immediate elastic deformation 
of the particles induced by the increase in shear stress. 

\begin{figure}[!htb]
 \begin{center}
 \resizebox{1. \columnwidth}{!}{%
\includegraphics{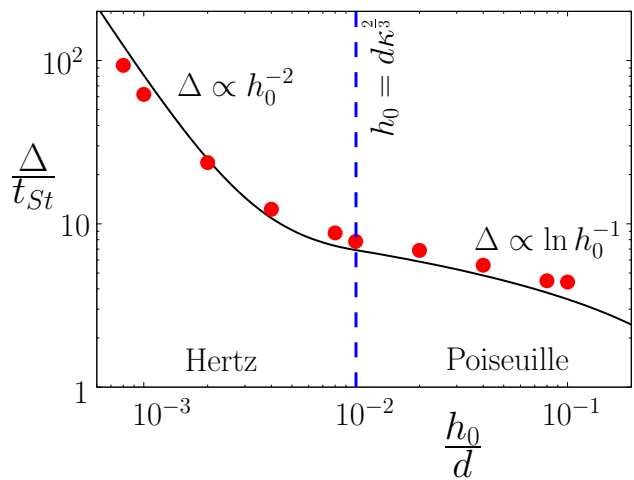}
 }
 \end{center}
 \caption{(Color online) Macroscopic relaxation time $\Delta$, 
estimated as shown on Fig.~\ref{Fig2}b,
as a function of the steric length $\gap_0$. 
Circles denote results from simulations,
while the curve corresponds to 3 times
the separation time $t^{sep}$ for two grains 
as defined by Eq.~(\ref{eq:tsep}).
The third data point ($\gap_0/d=2.10^{-3}$)
corresponds to curve F in Fig.~\ref{Fig4}.}
\label{Fig3}
\end{figure}

Fig.~\ref{Fig3} shows the value of $\Delta$ for materials with
different steric lengths $h_0$.
The creep duration $\Delta$ naturally increases 
when $h_0$ gets smaller.
However, as long as the steric length is larger than roughly $10^{-2}d$, 
$\Delta$ is small: only few times longer than the Stokes time.
In contrast, it increases strongly for smaller $h_0$. 

\subsection{Varying the step shear stress}

\begin{figure}[!htb]
 \begin{center}
 \resizebox{1.\columnwidth}{!}{%
 \includegraphics{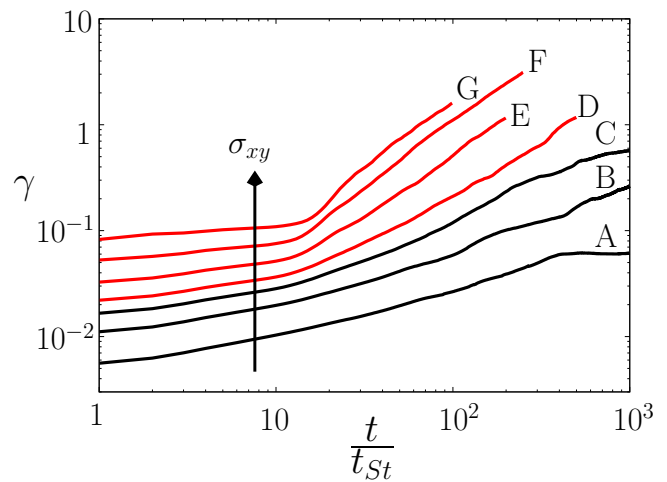}
 }
 \end{center}
 \caption{(Color online) Time evolution of the shear strain $\shear$ induced 
by establishing (from $t=0$ to $t=\tstokes$)
a constant shear stress (for larger $t$). 
Each curve corresponds to a different value
of the final shear stress:  $\sigD/\sigP = $ $0.05$
(A), $0.1$ (B), $0.15$ (C), $0.2$ (D), $0.3$
(E), $0.5$ (F) and $1.0$ (G). 
Red ({\em resp.} black) curves denote tests 
in which the material eventually flows
({\em resp.} does not flow).
The same material (with $h_0=2.10^{-3}$) was used for each run.
Curve F corresponds to one of the experiments
considered in Fig.~\ref{Fig3}.
}
\label{Fig4}
\end{figure}

In the second test, a material 
with a steric length $\gap_0=2.10^{-1}d$ 
was subjected to
varying step shear stress values (Fig.~\ref{Fig4}). 
It appears that below a critical
shear stress threshold, the material creeps 
and then does not seem to flow, but rather
reaches a static equilibrium within a finite shear strain 
smaller than 100\%. 
Above this shear stress threshold, 
here found to be roughly $\sigD = 0.2 \sigP$, 
the material creeps and then flows. 
One might imagine that the response delay
in the creep phase before the onset of flow
should be shorter if the applied constant shear stress
is higher.
Surprisingly, as can be seen from curves D--G of Fig.~\ref{Fig4},
{\em i.e.}, with applied shear stresses that differ by a factor of 5,
it seems that the duration of the creep phase 
remains essentially constant
although the transition from creep to flow
becomes sharper for larger values of the applied shear stress.

\section{Microscopic origin 
of the observed relaxation time} \label{sec:discussion}

Let us now interpret the macroscopic behaviors observed in the
previous section in terms of the dynamics 
of two particles in relative motion.
When the material is sheared, neighbouring particles need to move 
towards and away from each other. 
This relative motion includes sliding, 
rotation and normal approach and separation. 
As shown in Appendix~\ref{app:friction}, 
the normal viscous friction is the highest, 
which implies that this mode of motion should be the slowest. 
As a result, the material lies most of the time
in a situation where interparticle contacts
can be viewed as frictionless as far as sliding is concerned,
with only normal forces acting to bring particles
into contact or away from each other in a slow motion.
Hence, we focus on the sole normal mode 
for interpreting the macroscopic observations.

The duration of the normal separation process 
depends on the initial distance between
surfaces, on the viscous friction
and on the typical force transmitted by the fluid in the gap. 
The typical initial distance between
the surfaces can be estimated as the steric distance $\gap_0$. 
As the step in shear stress is applied 
to the dense material, 
the initial viscous friction depends on
whether the interaction is initially 
in the Poiseuille regime ($\delta<\gap_0$) 
or in the Hertz limit ($\delta>\gap_0$). 
Initially, the typical force transmitted within a pair of particles
scales with $\sigP d^2$. 
Hence, the interaction lies in the Hertz limit
when $\gap_0/d < \kappa^\frac{2}{3}$.

Let us assume that, when the shear stress is applied, 
the typical transmitted force is still of the order of $\sigP d^2$. 
Then, the time $t^{sep}$ for two particles to separate 
is such that $\int_{t=0}^{t^{sep}} \sigP d^2 dt
= \int_{h=h_0}^{h_{sep}} \frac{3}{2} \pi 
\viscofluid \frac{a^4}{\gap^3} d\gap $, 
where the dimension of the force-transmitting region is 
$a=\sqrt{dh}$ in the Poiseuille limit
and $a=\sqrt{d|\delta|}$ in the Hertz limit.
Hence: 

\begin{widetext}
\bee 
\frac{t^{sep}}{\tstokes} &=& \left\lbrace 
\begin{array}{l l}
\frac{1}{2} \ln{\left(\frac{h_{sep}}{h_{0}} \right)} & 
\mbox{(Poiseuille, $\frac{h_0}{d}\gg \kappa^{\frac{2}{3}}$)}, \\
\frac{1}{4} \left[ \kappa^{\frac{4}{3}} \left(\frac{d}{h_0} \right)^2
-1 \right] + \frac{1}{2} \ln{\left(\frac{h_{sep}}{d \kappa^{\frac{2}{3}}} \right)} 
&\mbox{(Hertz, $\frac{h_0}{d}\ll \kappa^{\frac{2}{3}})$}.\\ 
\end{array} \label{eq:tsep}
\right.
\eee
\end{widetext}
\noindent An arbitrary value of $h^{sep}$ has to be defined at the end
of the separation process. Basically, it should be of the order of $d$. But the time of
separation is only slightly depending on it. In the Poiseuille limit, the time of
separation
is at the order of the stokes time, with almost no effect by the small steric length
$\gap_0$. In
the Hertz limit, in contrast,
it is strongly dependent on $\gap_0$ 
as well as on the reduced normall stress
$\kappa 
= \frac{\sigP}{E}$. When applied to the system simulated in the previous section ($\kappa
= 10^{-3}$), this simple separation time $t^{sep}$ captures the scaling behavior of the
creep time $\Delta$ (Fig.\ref{Fig3}). The data are quantitatively
well represented  by the function $b t^{sep}$ involving a fit parameter $b \approx 3$.
This constant is likely to rely on the details of the
contact network (number of coordination, heterogeneities, anisotropy...) and may also
slightly depend on the geometrical definition of $\Delta$ (Fig.\ref{Fig2}b). The
salient result is that, surprisingly enough, this very simple model with only one
interaction and a single force scale $\sigP d^2$ captures the macroscopic delay observed:
both $t^{sep}$ and $\Delta$ scales with $\ln h_0^{-1}$ while the interaction is initially
in the Poiseuille regime ($h_0 > \kappa^\frac{2}{3} = 10^{-2}$), and with
$h_0^{-2}$ if the interaction is initially in the Hertz regime
(see Fig.~\ref{Fig3}). 

However, this model does not account for the secondary effect 
of the shear stress we observed in Fig.~\ref{Fig4}.
Basically, one could expect
that under a larger applied shear stress, 
the typical tensile pairwise normal force
would increase. 
Accordingly, one should observe a shorter separation time.
Fig.~\ref{Fig4} reveals that the creeping time
appears to be essentially independent of the applied shear stress.
This may have some connection,
although probably not direct,
with the counterintuitive effect 
highlighted in~\cite{Rognon08EPJE} 
concerning the crossover between the Poiseuille and the Hertz regimes: 
there exists an optimal tensile force that minimises 
the time of separation between two particles. 
This optimal force corresponds to the transition 
between the Hertz and the Poiseuille regimes.
Thus, it depends on the gap $\gap$ as: 
$F^{opt}\approx \young d^2 \left(\frac{h}{d}\right)^{\frac{3}{2}}$ 
(see~\cite{Rognon08EPJE} for more details). 
Increasing the force then brings the system 
further into the Hertz regime, 
which strongly increases the viscous friction
and eventually results in a slower separation.

\section{Conclusion}


The distinction between the flowing and non-flowing states 
of many materials is critical for determining failure. 
The current paper focuses on studying the micromechanical origins
of this phenomenon in relation to ideal materials 
of closed-packed elastic particles in a viscous fluid. 
Using {\em Soft Dynamics}
simulations, we show that when subjected to a step shear stress, this material can exhibit
a long creep phase before a steady flow state sets in. 
The typical creep time seems to rely on 
an internal relaxation process: 
the normal separation of two particles, 
which is driven by the applied stress and
resisted by the viscous friction. This internal relaxation time thus depends on 
{\em (i)} the applied stress, {\em (ii)} the particle stiffness and
{\em (iii)} the steric length $h_0$ below which
particle surfaces interact directly. 

The direct surface-to-surface interaction enables the
material to statically sustain a confining stress, and whether or not the steric
length is shorter than the typical elastic deflection strongly affects the scaling
behavior of the viscous friction. 
On one hand, the relaxation time is not very long in
the Poiseuille regime (large $h_0$ and/or very stiff particles) as its scales with
$\ln (h_0^{-1})$ and is thus only slightly longer than the Stokes time. 
On the other hand, the relaxation time can be very long 
in the Hertz limit (small $h_0$ and/or soft particles) 
as it then scales with $h_0^{-2}$ (see the estimation in table~\ref{tab:times}).

The long aging process corresponding to the Hertz limit should be a common characteristic
of various immersed particulate materials like granular pastes, emulsions and foams. We
expect that the soft bubbles and droplets would reach the Hertz regime for relatively low
confining stresses. For the solid grains, a rough estimate can be performed according to
Hertz's elasticity which defines the Hertz limit for $h_0 < d
\left(\frac{\sigP}{\young}\right)^{\frac{2}{3}}$. Assuming that the steric length
$\gap_0$ is determined by a surface roughness of amplitude $0.1\,\mu{\rm m}$, that
the grain size is $d\simeq 1\,{\rm mm}$ and that their Young modulus is $E\simeq
10^{10}\,{\rm Pa}$, then the normal stress required to reach the Hertz regime
would be on the order of $\sigP\simeq 10^4\,{\rm Pa}$. Note, however, that these materials
involve further modes of interaction such as solid
friction, interfacial rheology and adhesion, which would {\em (i)} induce other time
scales and {\em (ii)} affect the hydrodynamics within the gap and thus the expression of
the viscous friction which the aging process relies on. 
In future works, we hope to include some such detailed interactions 
within our methodology.

\begin{acknowledgements}
Financial supports for this research from the Australian Government's
Flagship Collaboration Fund through the CSIRO Flagship Cluster on Subsea
Pipelines and from the French
Government through the Agence Nationale de la Recherche (ANR-05-BLAN-0105-01) are
appreciated.
\end{acknowledgements}

\appendix

\section{Lubrication between elastic surfaces}\label{app:friction}

The viscous flow between between two close elastic spheres is not a simple
problem. 
However, simple scaling formulae can be deduced considering that only a small
portion
of the surface transmits the force (size $a$), and that its radius of
curvature is large ($a \ll R$). Then, the hydrodynamical interaction is similar to that
which develops between two flat parallel disks of radius $a$ separated by $h$. The low
aspect ratio of
the gap, $\frac{h}{a}\ll 1$ favors flows in the Stokes regime.

\begin{figure*}[!htb]
 \begin{center}
 \resizebox{2.\columnwidth}{!}{%
 \includegraphics{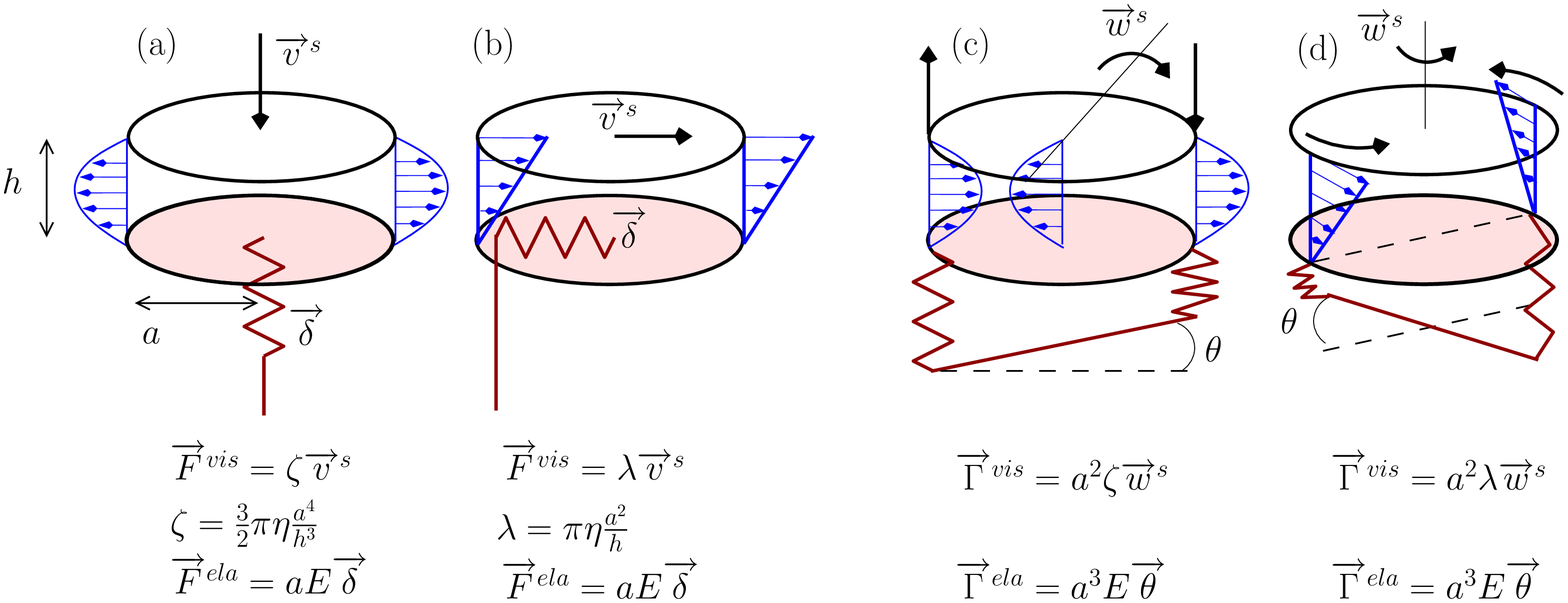}
 }
 \end{center}
 \caption{ (Color online) Scheme of the elasto-hydrodynamical interaction implemented in
{\em Soft Dynamics}, 
including four modes of relative motion between particles with a
surface-to-surface distance
of $h$. Distance $a$ is the size of the transmitting force region (see Fig.\ref{Fig1}a). The blue
arrows designate the flow velocity profile 
and the red spring the elastic deformation of the bulk
material. The corresponding expression of viscous and elastic forces/torques are as
follows. (a) The
normal approach induces a Poiseuille flow and an associated viscous friction
$\zeta$ (also valid for a normal separation). (b) The tangential sliding induces a Couette
flow and an associated viscous friction $\lambda$. (c) Normal
rotation (locally, the flow is like a Poiseuille flow). (d) Parallel rotation (locally,
the flow is like a Couette flow).
}
\label{Fig5}
\end{figure*}

A simple way to get the scaling formulae for the viscous force and torque is to consider
the average shear strain rate $\dot \epsilon$ in the liquid, which scales like
$\frac{v^s}{h}$ for
a Couette flow and like  $ \frac{3}{2}\frac{v^s}{h} \frac{\pi a^2}{2\pi a\,h}$ for a
Poiseuille flow. On the one hand, the power dissipated by the viscous flow in a gap of a
volume $\pi a^2 \gap$ scales with $\eta {\dot\epsilon}^2\,a^2\gap$. On the other hand, this
power is equal to $F^{vis}v^s$ for a relative velocity $v^s$ and to $\Gamma^{vis}w^s$ for
a relative angular velocity $w^s$. This leads to the expressions of the viscous forces and
torques in Fig.~\ref{Fig5}. Note that the viscous force and torque can be expressed
through the viscous friction tensors $Z$ and $w$:

\be
\begin{array}{l l}
 \V{F}^{vis} = Z \cdot \V{v}^s  &\; ~with~ Z = \zeta \V{n} \otimes  \V{n} + \lambda
(I-\V{n} \otimes  \V{n})\\
\V{\Gamma}^{vis}  = W \cdot \V{w}^s & \; ~with~ W =a^2 Z,\\
\end{array}\label{eq:viscous_fG}
\ee

\noindent where $\V{n}$ is the unit vector normal to the surface, $I$ the unity tensor
and $\otimes$ the standard cross product.

The elastic forces and torques can also be estimated in a simple way by following the
Hertz
theory: the elastic strain $\epsilon$ concerns mostly a small region of volume
$a^3$. It scales according to $\frac{\delta}{a}$ for both the normal and the tangential
deflections represented by
$\delta$, and as $\frac{a\theta}{a}$ for a twist angle $\theta$. The total elastic
energy, $E\epsilon^2$, therefore scales like $Ea^3\left(\frac{\delta}{a}\right)^2$ for
deflections and according to $Ea^3 \theta^2$ for twists. The differentiation of these
energies with
respect to $\delta$ and $\theta$ gives the scaling for the elastic forces and torques
expressed in Fig.~\ref{Fig5}, where all numerical constants are set to unity for the sake of simplicity.

\section{Soft Dynamics method}\label{sec:motion}

We have developed the {\em Soft Dynamics} method to simulate large samples of closely packed
particles with elasto-lubricated interaction~\cite{Rognon08EPJE,
Rognon09EPJE}. 
In the present work, the particles are inertial so that their
motion is deduced from the forces and torques they
experience according to Newton's law, as in the standard discrete element
methods:

\be
\left\lbrace 
\begin{array}{l}
m_i \ddot{\V{ X}_i} =  \V{F}^{ext}_i + \sum_j \V{F}^{ela}_{ij}\\
J_i \dot{\V{\Omega}_i} = \V{\Gamma}^{ext}_i + \sum_j \left\lbrace
\V{\Gamma}^{ela}_{ij} + \V{r}_{ij}  \times \V{F}^{ela}_{ij} \right\rbrace .
\end{array}
\right.
\ee

\noindent $m_i$ and $J_i$ are respectively the mass and the inertia moment of particle
$i$, $\V {X}_i$ and $\V {\Omega}_i$ its position and its rotation velocity.
$\V{F}^{ext}_i$ and $\V{\Gamma}^{ext}_i$ are possible external force and torque that
the particle experiences, which are both taken as zero in the present study. The sums run on each
particle $j$ interacting with $i$. The force $\V{F}^{ela}_{ij}$ and the torque
$\V{\Gamma}^{ela}_{ij}$ come from the elastic deflection and twist of the particle
surfaces, and $\V{r}_{ij}$ is the vector from the center of $i$ to the interaction point.

The uniqueness of the {\em Soft Dynamics} approach 
is to include the dynamics of the elastic deflection
$\V{\delta}_{ij}$ and twist angle $\V{\theta}_{ij}$ 
so that the force and torque balances 
are satisfied for each interaction. $ \dot{\V{\delta }_{ij}}$ and
$\dot {\V{\theta}_{ij}}$ can be deduced from the difference between the particle
velocities, $\dot{\V{ X}_i}$ and $\V{\Omega}_i$, and the relative translational and
rotational velocities of
particle surfaces, $\V{v}^s_{ij}$ and $\V{w}^s_{ij}$. Compatibility then requires to have:

\be
\left\lbrace 
\begin{array}{l}
\dot{\V{\delta}_{ij}}  = \dot{\V{ X}_j} +  \dot{\V{\Omega}_j} \times \V{r}_{ji} - 
\left[ \dot{\V{ X}_i} +  \dot{\V{\Omega}_i} \times \V{r}_{ij} \right] -\V{v}^s_{ij} \\
\dot {\V{\theta}_{ij}}=  \dot{\V{\Omega}_j} - \dot{\V{\Omega}_i} - \V{w}^s_{ij} \\
\end{array}\label{eq:contact_motion}
\right.
\ee

\noindent For each of the interactions, the relative velocities $\V{v}^s_{ij}$ and
$\V{w}^s_{ij}$
are given by the force/torque balances:

\be
\left\lbrace 
\begin{array}{l}
\V{F}^{ela}_{ij}  =  \V{F}^{vis}_{ij} +\V{F}^{dir}_{ij} \\
\V{\Gamma}^{ela}_{ij}  =  \V{\Gamma}^{vis}_{ij} + \V{\Gamma}^{dir}_{ij}. \\
\end{array}\label{eq:contact_balance}
\right.
\ee

\noindent and the fact that the viscous force/torque are proportional to them through the
friction tensors $Z_{ij}$ and $W_{ij}$ (see App.~\ref{app:friction}):

\be
\left\lbrace 
\begin{array}{l}
\V{v}^s_{ij}  = Z_{ij}^{-1} \cdot \left[ \V{F}^{ela}_{ij} - \V{F}^{dir}_{ij} \right] \\
\V{w}^s_{ij}  = W_{ij}^{-1} \cdot \left[  \V{\Gamma}^{ela}_{ij}
- \V{\Gamma}^{dir}_{ij}\right] . \\
\end{array}\label{eq:contact_velo}
\right.
\ee

The second order equations (\ref{eq:contact_motion}) are integrated using a standard
predictor-corrector scheme, while the first order equations (\ref{eq:contact_balance})
are integrated using a simple Euler scheme. The time step of integration is a
fraction of the shortest time scale in the system, namely $t_{M}=\frac{\viscofluid}{\young}$.
Through careful examination we found that a time step of five time smaller than $t_{M}$ is
short enough.

Since we do not include long range hydrodynamical interactions, it is sensible to
introduce a cut-off
distance: here two particle interact only if their surface to surface distance is smaller
than
$0.25d$. The size of the transmitting force region $a$ is then being interpolated between
the two asymptotic
limits of the Poiseuille and the Hertz models, as a way to get a continuous transition. 
The interpolation we choose is: $a= \sqrt{d\left(h+|\delta| \right)}$. Note that
different choices of interpolation are possible, 
although this should have no incidence
except in the vicinity of the crossover
between both scaling regimes.


\end{document}